\documentclass[twocolumn,preprintnumbers,amsmath,amssymb]{revtex4}
\usepackage{graphicx}
\usepackage[dvipsnames]{xcolor}
\usepackage{ulem}
\newcommand{\be}{\begin{equation}}
\newcommand{\ee}{\end{equation}}
\newcommand{\beq}{\begin{equation}}
\newcommand{\eeq}{\end{equation}}
\newcommand{\bea}{\begin{eqnarray}}
\newcommand{\eea}{\end{eqnarray}}
\newcommand{\eml}{\end{mathletters}}

\begin{document}

\title{Bridging the quartet and pair pictures of isovector proton-neutron pairing }

\author{V.V. Baran$^{1,2}$ }
\email[Corresponding author: ]{vvbaran@fizica.unibuc.ro}
\author{ D. R. Nichita$^{1,2}$}
\author{D. Negrea$^{2}$}
\author{D. S. Delion$^{2}$ }
\author{N. Sandulescu$^{2}$ }
\author{P. Schuck$^{3}$ }
\affiliation{
$^1$ Faculty of Physics, University of Bucharest,
405 Atomi\c stilor, POB MG-11, Bucharest-M\u agurele, RO-077125, Romania\\
$^2$"Horia Hulubei" National Institute of Physics and 
Nuclear Engineering, \\ 
30 Reactorului, RO-077125, Bucharest-M\u agurele, Romania \\
$^3$ Universit\' e Paris-Saclay, CNRS, IJCLab, IN2P3-CNRS, 91405 Orsay, France\\
Universit\' e Grenoble Alpes, CNRS, LPMMC, 38000 Grenoble, France
}

\begin{abstract}

The formal implications of a quartet coherent state ansatz for proton-neutron pairing are analyzed. Its nonlinear annihilation operators, which generalize the BCS linear quasiparticle operators, are computed in the quartetting case. Their structure is found to generate nontrivial relationships between the many body correlation functions. The intrinsic structure of the quartet coherent state is detailed, as it hints to the precise correspondence between the quartetting picture and the symmetry restored pair condensate picture for the proton-neutron pairing correlations.

\end{abstract}

\maketitle

\subparagraph{Introduction.}

After more than sixty years since pairing effects were first considered in nuclear physics \cite{Bohr58}, the microscopic pairing models are still facing the challenge of consistently describing the subtle interplay between the isovector (T=1) and the isoscalar (T=0) proton-neutron pairing in nuclear systems \cite{Fra14}. 
One of the first studies of the isovector pairing Hamiltonian was performed by
Beliaev et al in the framework of the generalised BCS approach in which the protons and the neutrons
are mixed through the Bogoliubov transformation \cite{Bel60}. Since then, the BCS approach was
employed in the majority of studies and it was further extended to include also the isoscalar
proton-neutron pairing interaction \cite{Goo79}. However, as has been noticed already by  Beliaev
et al,  the BCS treatment is not complete because “one must take into consideration the
quadruple correlation of $\alpha$ particle-like nucleons in addition to pair correlations”. The first
investigation of these correlations has been done by Soloviev \cite{Sol60}, who related them to a
4-body interaction term. Later on, the 4-body ``quartet" correlations have been discussed in
relation to the standard two-body isovector pairing interaction by Br\'emond and Valatin \cite{Bre63}
and by Flowers and Vujicic \cite{Flo63}. They proposed a BCS-like function in which the pairs are
replaced by quartets, but the calculations with this trial state turned out to be too complicated
and it was never applied to realistic cases. The first proof that the quartets are essential
degrees of freedom for the isovector pairing Hamiltonian was given by Dobes and Pittel \cite{Dob98}
for the particular case of degenerate shells. They have shown that in this case the exact solution 
of the isovector pairing Hamiltonian for even-even N=Z systems can be expressed as a quartet condensate, 
with the quartet defined as two isovector pairs coupled to total isospin T = 0. Later on quartet condensation 
models (QCM) have been proposed for non-degenerate levels and applied for realistic isovector pairing 
Hamiltonians \cite{San12,San12a}. Recently it was shown that the exact solution for the non-degenerate levels
can be also expressed in terms of quartets \cite{Samb20a} and that this solution turns to a quartet 
condensate in the strong coupling coupling \cite{Samb20b}. All these studies have demonstrated that the
$\alpha$-like quartets are indispensable for a proper description of isovector pairing.

At this point it is worth stressing that the quartet condensation in the pairing context
mentioned above should not be confused with the other `quartet condensate' concept, based
on a similar wave function as the QCM one but dealing with {\it in medium bound states} of 
four fermions as, e.g., alpha particles, and their Bose-Einstein
condensation \cite{Sogo09, THSR} in finite nuclei and infinite nuclear matter. While no alpha particle condensate survives at  saturation density, one may develop a theory for quartet condensation which in many aspects is similar to the BCS approach for the condensation of pairs \cite{Schuck2014}.

Very recently we introduced such a BCS-like approach involving a quartet coherent state in the context of proton-neutron pairing \cite{QBCS}.
The aim of this work is to further explore the implications of a quartet coherent state ansatz for the proton-neutron pairing problem. This leads us to establish the general relation between the quartet models and the BCS-based models, which was previously investigated only for particular cases \cite{Dob98,San12,San12a,San09,Dob19}.

We consider the general isovector pairing Hamiltonian
\beq
\label{ham}
H=\sum_{i=1}^{N_\text{lev}}\epsilon_i N_{i,0}+\sum_{\tau=0,\pm1}\sum_{i,j=1}^{N_\text{lev}}V_{ij}P^{\dagger}_{i,\tau } P_{ j,\tau}~,
\eeq
where $i,j$ denote the single particle four-fold degenerate states and $\epsilon_i$ refers to the single particle energies; a time conjugated state will be denoted by $\bar{i}$. The first part is the standard single-particle term while the second
part is the isovector pairing interaction expressed by the neutron-neutron ($\tau=1$), proton-proton
($\tau=-1)$ and proton-neutron ($\tau=0$) pairs operators defined by $P^\dagger_{i,\tau}=[c^\dagger_{i}c^\dagger_{\bar{i}}]^{T=1}_\tau$. In the discussion below, we will frequently refer to the set of collective $\pi\pi$, $\nu\nu$ and $\pi\nu$ Cooper pairs $\Gamma^\dagger_\tau(x)\equiv\sum_{i=1}^{N_\text{lev}}x_i P^{\dagger}_{i,\tau}$, 
which depend on a set of mixing amplitudes $x_i$, $i=1,2,...,N_{\text{lev}}$. We denote by $
q_i^\dagger=\nu^\dagger_{i,\uparrow}\nu^\dagger_{i,\downarrow}\pi^\dagger_{i,\uparrow}\pi^\dagger_{i,\downarrow}$ the isoscalar quartet operator that fills completely the level $i$.

The BCS-like quartet coherent state ansatz introduced in Ref. \cite{QBCS} is written in terms of the QCM collective quartet operator $Q^\dagger(x)\equiv \sqrt{3}\left[\Gamma^\dagger\Gamma^\dagger\right]^{T=0}\equiv 2\Gamma^\dagger_1(x)\Gamma^\dagger_{-1}(x)-[
\Gamma^\dagger_0(x)]^2$ as
\bea
|QBCS\rangle&=&\exp[Q^\dagger]\,|0\rangle=\sum_n \frac{1}{n!}[Q^\dagger]^n|\,0\rangle~.
\eea
Below, we shall explore in more detail the particular consequences of its coherent state character.

The paper concentrates on formal aspects of pairing and quartetting. We will first show that QBCS of Eq. (2) can be annihilated by a non-linear transformation of fermion operators (mixing singles and triples and/or doubles with quadruples). This is in analogy to the well-known, simpler case where a quasi-particle operator annihilates the BCS state. We will discuss how the former can open very interesting possibilities of calculus  with the quartet coherent states. An interesting aspect will be that QBCS can be written as a Hubbard-Stratonovich transformation of a single particle field. This will help to show that the number projected QBCS is analytically equivalent to the number projected BCS for $T=0$ states. For $T > 0$ states the equivalence will be shown only numerically, getting very close to 100$\%$.

\subparagraph{QBCS annihilation operators.} One of the major advantages of the BCS approach is the possibility of describing the paired system in a picture of weakly interacting ``quasiparticles'', whose associated operators obey an annihilation condition with respect to the correlated BCS vacuum. Despite its nonlinear character, the above quartet-BCS state still admits a generalized class of annihilation operators, due to its coherent state nature \cite{Schuck2014}.  However, at variance with the linear quasiparticles of the BCS case, the annihilation operators in the quartetting case do not obey simple linear equations of motion.  For a specific particle operator $c$ and for a specific pair operator $P\sim c\, c$, the general annihilation operators may be computed as
\beq
\label{killers}
\begin{aligned}
\alpha&= c +[Q^\dagger,c]~,\\
\beta&=P+[Q^\dagger,P]_{(2)}+\frac{1}{2}[Q^\dagger,P]_{(3,1)}~,
\end{aligned}
\eeq
where we used the decomposition $[Q^\dagger,P]\equiv[Q^\dagger,P]_{(2)}+[Q^\dagger,P]_{(3,1)}$ which is of the form $c^\dagger c^\dagger + c^\dagger c^\dagger c^\dagger c$.
Explicitly, a proton-like annihilation operator of the QBCS state has the form
\beq
\label{pkiller}
\alpha_{i,\uparrow}= \pi_{i,\uparrow} -2\,x_i\,\pi^\dagger_{i,\downarrow}\Gamma^{\dagger}_{1}(x)+\sqrt{2}\,x_i\,\nu^\dagger_{i,\downarrow}\Gamma^{\dagger}_{0}(x)~,
\eeq
involving the annihilation of a particle and the creation of a particle dressed by a collective pair. Analogous relations hold for the other spin-isospin combinations.

The specific form of these nonlinear annihilation operators has interesting consequences; one in particular is the existence of a nontrivial connection between the two-body normal densities and the four-body anomalous densities. To see this, evaluate the average $\langle QBCS|\pi^\dagger_{i,\uparrow} \alpha_{i,\uparrow} +\pi^\dagger_{i,\downarrow} \alpha_{i,\downarrow} |QBCS\rangle=0$ by using Eq. (\ref{pkiller}). This leads to the occupations expressed in terms of the quartetting tensor as
\beq
\label{nz}
\begin{aligned}
n_i&= \langle \pi^\dagger_{i,\uparrow}\pi_{i,\uparrow}+\pi^\dagger_{i,\downarrow}\pi_{i,\downarrow}\rangle=2\,x_i\, \sum_j x_j \langle [P^\dagger_i P^\dagger_j]^{T=0}\rangle~,
\end{aligned}
\eeq
where the averaging is performed on the QBCS state. It follows that the total number of quartets may be expressed as the average of the collective quartet operator as
\beq
n_q =\langle QBCS|\, Q^\dagger(x)\,|QBCS\rangle~.
\eeq
This is a generalization of the simple BCS case with the ground state $|BCS\rangle \sim \exp[{\Gamma^\dagger(x)}]|0\rangle$ and the annihilation operators $\alpha_{i,\uparrow}=c_{i,\uparrow}-x_ic^\dagger_{i,\downarrow}$. Here the occupations may be computed  from $0=\langle c^\dagger_{i,\uparrow}\alpha_{i,\uparrow}\rangle=n_i/2-x_i\langle c^\dagger_{i,\uparrow}c^\dagger_{i,\downarrow}\rangle$. It follows that the number of pairs is given by the average of the collective pair operator 
\beq
n_p=\sum_i n_i/2=\langle BCS| \Gamma^\dagger(x)|BCS\rangle~. 
\eeq
For the BCS case, we may also introduce the occupation and unoccupation amplitudes and recover the familiar form $n_p=\sum_i x_i \langle c_{i,\uparrow}^\dagger c_{i,\downarrow}^\dagger\rangle=\sum_i (v_i/u_i) u_iv_i=\sum_i v_i^2$.

Returning to the second class of pair-like QBCS annihilation operators, the  expressions resulting from Eq. (\ref{killers}), for each isovector pair $P_{k,\tau}$, are
\beq
\label{killers1}
\begin{aligned}
\beta_{k,\pm1}&=P_{k,\pm1}-x_k^2P_{k,\mp1}^\dagger-2x_k \Gamma_{\mp1}^\dagger\\
&+x_k\Gamma_{\mp1}^\dagger N_{k,\pm1}-x_k \Gamma_0^\dagger T_{k,\mp1}~,\\
\beta_{k,0}&=P_{k,0}+x_k^2P_{k,0}^\dagger+2x_k \Gamma_0^\dagger\\
&+x_k\Gamma_1^\dagger T_{k,-1}-x_k\Gamma_{-1}^\dagger T_{k,1}-\frac{1}{2}x_k \Gamma_0^\dagger N_{k,0}~,\\
\end{aligned}
\eeq
involving pair creation and annihilation terms, together with a nonlinear pair dressed by the particle number and isospin operators, $T_{i,1} = - (  \pi^{\dagger}_{{i},\uparrow} \nu_{{i},\uparrow} +\pi^{\dagger}_{i,\downarrow} \nu_{i,\downarrow})/\sqrt{2}$ and $T_{i,-1} = -T_{i,1}^\dagger$. Remarkably, there is another nonlinear combination that commutes exactly with the quartet operator. Explicitly, with
\beq
\eta_k\equiv \Gamma_1^\dagger T_{k,-1}-\Gamma_{-1}^\dagger T_{k,1}+\frac{1}{2} \Gamma_0^\dagger N_{k,0}-\frac{1}{2}x_kP^{\dagger}_{k,0}\sum_{j}N_{j,0}~,
\eeq
we have $[Q^\dagger,\eta]=0$ and thus $\eta_k|QBCS\rangle=0$. Because the isospin operators, $T_{\pm1}=\sum_k T_{k,\pm1}$, obey   $[Q^\dagger,T_{\pm1}]=0$, they also annihilate the isospin conserving QBCS state, $T_{\pm1}|QBCS\rangle=0$. 

The annihilation of the QBCS state by the operators $\eta_k$ and $T_{\pm 1}$ leads to the fact that the operators in Eq. (\ref{killers1}) are not actually uniquely defined. We could add to any of the $\beta$'s an arbitrary combination of $\eta$ and $P T$ and still obtain a valid pair-like annihilation operator. This freedom could allow for new treatments to be consistently developed for the pairing Hamiltonian, in analogy with Refs. \cite{Schuck2014, Schuck2020}, as will be explored in future works.

\subparagraph{Structure of the QBCS state.} Computations with the nonlinear QBCS ansatz are made tractable in Ref. \cite{QBCS}  by a linearization procedure for the exponent. The quartet operator is first expressed as the square of a rotated collective pair $\gamma$, $Q=\vec{\gamma}^{\dagger}\cdot \vec{\gamma}^\dagger$,
 defined by $\gamma_\tau^\dagger=\sum_{j=1}^{N_\text{lev}} x_j p_{j,\tau}^\dagger$, where
\beq
\label{rotpairs}
\begin{aligned}
p_{j,1}^\dagger&=i({P_{j,1}^\dagger-P_{j,-1}^\dagger})/{\sqrt{2}},~ p_{j,2}^\dagger=({P_{j,1}^\dagger+P_{j,-1}^\dagger})/{\sqrt{2}}~,\\
 p_{j,3}^\dagger&=-i P_{j,0}^\dagger~.
\end{aligned}
\eeq
Note that this choice is not unique. A Hubbard-Stratonovich transformation is then used to represent the quartet coherent state as a combination of general isovector pair BCS states,
\beq
\label{qbcs1}
\begin{aligned}
&\exp(Q^\dagger)=\exp(\vec{\gamma}^{\dagger}\cdot \vec{\gamma}^\dagger)= \\
&\int\text{d}^3z\exp\left(-\vec{z}^{\,2}/4+\vec{z}\cdot\vec{\gamma}^\dagger \right)=\\
&\int\text{d}^3z\, e^{-\vec{z}^{\,2}/4 } \prod_{i=1}^{N_{\text{lev}}} (1+x_i \vec{z}\cdot \vec{p}^{\, \dagger}_i+x_i^2 \vec{z}^{\,2}q^\dagger_i/2)~,
\end{aligned}
\eeq
where we omitted the overall normalization factor. 
In this way, we obtain a superposition of standard BCS states, each factorized as a product over the single particle levels.

To better understand this specific pattern of partial symmetry breaking, it is instructive to pass to spherical coordinates in Eq. (\ref{qbcs1}) and write the quartet coherent state as
\beq
\label{qbcs2}
\begin{aligned}
\exp(Q^\dagger)&=\int_0^\infty \text{d}z\, z^2 e^{-{z}^{2}/4 }\int_{S^2}\text{d}\hat{n}\, \exp(z\, \hat{n}\cdot \vec{\gamma}^{\, \dagger})~.
\end{aligned}
\eeq
Naturally, the isospin projection is already implemented by the angular integration. To see this, consider the coherent state of the isovector pair $\vec{\gamma}$ integrated over all directions in isospace, 
\beq
\label{qbcs3}
\begin{aligned}
j_0^\dagger&\equiv\int_{S^2}\text{d}\hat{n}\, \exp( \hat{n}\cdot \vec{\gamma}^{\, \dagger})\\
&=\sum_{k=0}^\infty\frac{(\vec{\gamma}^\dagger\cdot \vec{\gamma}^\dagger)^k}{(2k+1)!}=\sum_{k=0}^\infty\frac{(Q^\dagger)^k}{(2k+1)!}=j_0(i\sqrt{Q^\dagger})~,
\end{aligned}
\eeq
which is formally the expansion of a spherical Bessel function of imaginary argument (hence the name). The basic information about the quartet correlations is thus already contained in this simpler ansatz; by projecting onto good particle number, we always recover the QCM state,
\beq
\mathcal{P}_{n_q} \exp(Q^\dagger)\, |0\rangle=\mathcal{P}_{n_q}\, j_0^\dagger \, |0\rangle=(Q^\dagger)^{n_q}\,|0\rangle~.
\eeq
We interpret now the role of the radial integral in Eq. (\ref{qbcs2}) as just changing the mixing between the components having different particle numbers. 

The analytic expressions of the norm function and of the Hamiltonian average on the $j_0^\dagger$ state may be obtained simply by dropping the radial integrals from the QBCS expressions (see Ref. \cite{QBCS}, Supplemental Material). Remarkably, identical expressions were reported in Refs. \cite{Kyotoku79,Kyotoku87}, in the context of the symmetry restored BCS approach. The definition itself of the $j_0^\dagger$ state hints at a precise relationship with the projected BCS state, which we detail below.

\subparagraph {BCS Symmetry restoration for $T=0$.}

The generalised BCS equations for isovector pairing in even-even $N=Z$ systems present two degenerate solutions with gap parameters $\Delta_\nu=\Delta_\pi=\Delta, \Delta_{\pi\nu}=0$, and $\Delta_\nu=\Delta_\pi=0, \Delta_{\pi\nu}=\Delta$ (for a proof, see \cite{San09}). The corresponding BCS states are given by
\beq
\begin{aligned}
|BCS_I\rangle&= \exp[\Gamma^\dagger_{1}(x)]\,\exp[\Gamma^\dagger_{-1}(x)]\, |0\rangle~,\\
\label{bcs2}
|BCS_{II}\rangle&= \exp[\Gamma^\dagger_{0}(x)]\,|0\rangle~.
\end{aligned}
\eeq

Techniques for projecting these solutions
onto good particle number and isospin have been developed in \cite{Kyotoku79,Kyotoku87,Chen78, Raduta00,Raduta01,Raduta00a,Raduta12}, with their connection to the quartet models only being mentioned {for particular cases} in Refs. \cite{Dob98,San09,Dob19}. 

Here, we establish the correspondence in the general case by analytically performing the projection operation on the BCS state, and recovering a version of the $j_0^\dagger$ ansatz of Eq. (\ref{qbcs3}).  For simplicity, we consider the axially symmetric state $|BCS_{II}\rangle$ with $T_z=0$ and we  employ the isospin projection operator \cite{RS}
\beq
\mathcal{P}_{T;T_z=0}= \int_{S^2} \text{d}\hat{n}\, D^{T*}_{0 0}(\hat{n})\, R(\hat{n})~,
\eeq
written in terms of a Wigner $D$-matrix and of the rotation operator in isospin space   $R(\hat{n})$, which may be factorized as $ R(\hat{n}) =\prod_{i=1}^{N_{\text{lev}}}R_i(\hat{n})$. Given the isoscalar character of the fully occupied single particle level $q^\dagger_i|0\rangle$, the only nontrivial term involves the rotation of the one-pair state.  The isospin rotation operator $R_i(\hat{n})=\exp(-i\, \varphi\,  \hat{T}_z)  \exp(-i\, \theta\,  \hat{T}_y)$ acting on a $T_z=0$ pair state is effectively
\beq
\begin{aligned}
R_i(\hat{n})\, P^{\dagger}_{i,0}\, R_i(\hat{n})^{-1}=i\, \hat{n}\cdot \vec{p}^{\,\dagger}_i~,
\end{aligned}
\eeq
involving the same rotated pairs $\vec{p}^{\,\dagger}_i$ of Eq. (\ref{rotpairs}) used to bring the collective quartet operator to a diagonal form. The isospin rotated BCS state becomes
\beq
\begin{aligned}
 R(\hat{n}) |BCS_{II}\rangle &=\prod_{k=1}^{N_{\text{lev}}}(1+i\, x_k\, \hat{n}\cdot \vec{p}^{\,\dagger}_k-x_k^2\,q^\dagger_k/2)|0\rangle\\
 &=\exp(i\, \hat{n}\cdot \vec{\gamma}^{\,\dagger})~.
\end{aligned}
\eeq
This implies that the isospin projected BCS may be written as
\beq
\mathcal{P}_{T;T_z=0} |BCS_{II}\rangle= \int_{S^2} \text{d}\hat{n}\, D^{T*}_{0 0}(\hat{n})\,\exp(i\, \hat{n}\cdot \vec{\gamma}^{\,\dagger})~.
\eeq
In particular, the $T=0$ component is simply
\beq
\begin{aligned}
\mathcal{P}&_{T=0} |BCS_{II}\rangle= \int_{S^2} \text{d}\hat{n}\,\exp(i\, \hat{n}\cdot \vec{\gamma}^{\,\dagger})\\
&=\sum_{k=0}^\infty\frac{(-\vec{\gamma}^\dagger\cdot \vec{\gamma}^\dagger)^k}{(2k+1)!}=\sum_{k=0}^\infty\frac{(-Q^\dagger)^k}{(2k+1)!}=j_0(\sqrt{Q^\dagger})~,
\end{aligned}
\eeq
which is nothing else than  Eq. (\ref{qbcs3}) evaluated with imaginary mixing amplitudes or, equivalently, originating from the ansatz $\exp(-Q^\dagger$). 

This proves the general equivalence of the projected BCS and QCM approaches, for the isovector pairing correlations in the $T=0$ ground state of $N=Z$ even-even nuclei, i.e.
\beq
\mathcal{P}^{\, \mathcal{N}=4n_q}_{T=0}|BCS\rangle=(Q^\dagger)^{n_q}|0\rangle=|QCM\rangle~.
\eeq 

Before detailing with the $N>Z$ case below, we   remark the possibility of establishing nontrivial connections between the correlation functions also for the particle number projected $QCM$ state, based on the above annihilation operators. We write Eq. (\ref{pkiller}) in schematic form $\alpha=c+c^\dagger c^\dagger c^\dagger$, and project the annihilation condition $\alpha \exp[Q^\dagger]|0\rangle=0$ onto a fixed particle number, which singles out two terms. A proper particle-like annihilation operator for the $QCM$ state may then be expressed in terms of the inverse amplitude coherent quartet, which satisfies $
Q(1/x)\, Q^{\dagger}(x) |QCM\rangle=\lambda |QCM\rangle$, with $\lambda$ a numerical factor (for details see Appendix A of \cite{phb}). We obtain e.g., for the proton-like annihilation operator,
\beq
\label{QCMkiller}
\left[\pi_{i,\uparrow}+\frac{n_q}{\lambda}[Q^\dagger,\pi_{i,\uparrow}] \, Q\left(\frac{1}{x}\right)\right]|QCM\rangle=0
\eeq
where the commutator can be read off Eq. (\ref{pkiller}).
In analogy with Eq. (\ref{nz}) for the quartet coherent state, we may obtain a relation between the particle and the quartet densities on the $QCM$ state of the form $\langle QCM |c^\dagger c|QCM\rangle=\langle QCM |c^\dagger c^\dagger c^\dagger c^\dagger c  c c c|QCM\rangle$. 

This is perfectly analogous to the simple single-species BCS case, where the quasiparticle action on the BCS state $(c_{i,\uparrow}-x_i c^\dagger_{i,\downarrow})\exp[\Gamma^\dagger(x)]|0\rangle=0$ may be projected to obtain the nonlinear annihilation relation
\beq
\left[c_{i,\uparrow}-\frac{x_i}{N_{\text{lev}}-n+1} c_{i,\downarrow}^\dagger \Gamma\left(\frac{1}{x}\right)\right][\Gamma^\dagger(x)]^n|0 \rangle=0
\eeq
We may then find the connection between the particle and the pair densities on the projected BCS state $|PBCS\rangle=[\Gamma^\dagger(x)]^n|0 \rangle$ as
\beq
\langle c^\dagger_{i,\uparrow} c_{i,\uparrow}\rangle=\frac{x_i}{N_{\text{lev}}-n+1}\sum_{j=1}^{N_{\text{lev}}}\frac{1}{x_j}\langle P^\dagger_i P_j \rangle
\eeq
Similar relationships may be established also for higher order correlation functions, which could enable new ways of solving the pairing problem, e.g. within the recent many body bootstrap approach \cite{qmb_boot, boot_prl}.

\subparagraph{QCM vs projected BCS for $N>Z$.}

\begin{figure*}[ht]

 \centering

\includegraphics[width=\textwidth]{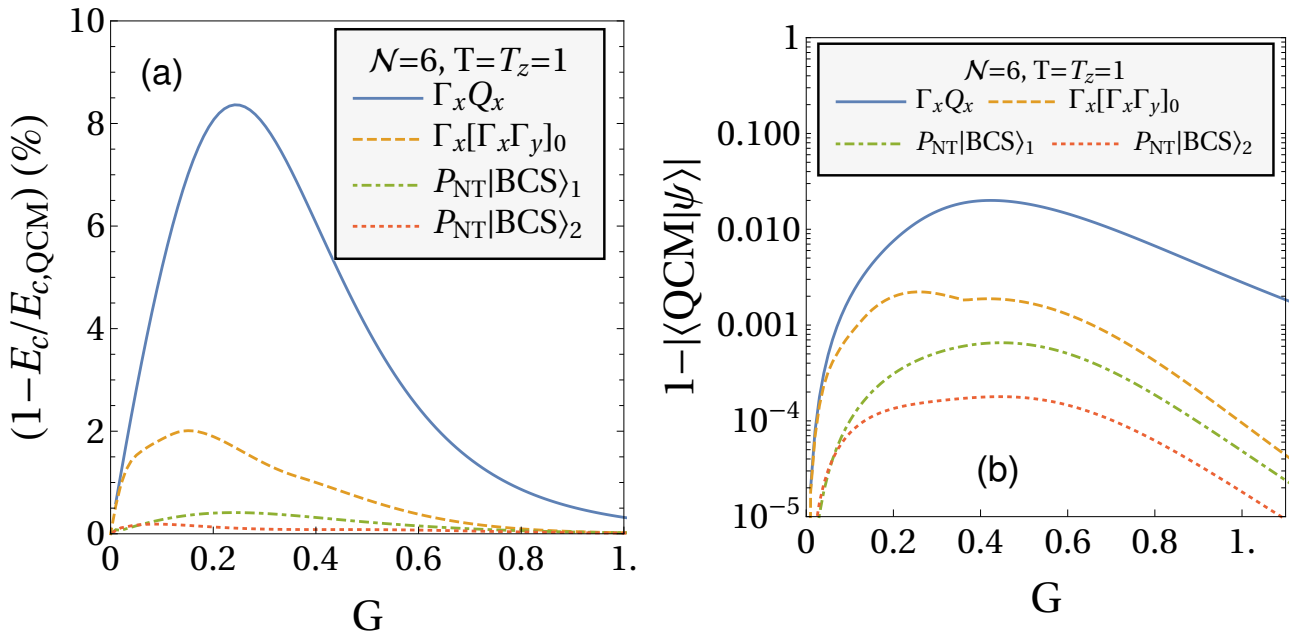}

\caption{Error in the correlation energy $E_c=E(G)-E(G=0)$ (a) and overlaps (b) {for the states (26) relative to the QCM state (25)}, versus the interaction strength G (in units of the level spacing). {The results are for two protons and four neutrons on eight equidistant levels. To indicate the
results we use the notations:} ``$\Gamma_x Q_x$" for the state $\Gamma^\dagger(x)Q^\dagger(x)|0\rangle$ of Eq. (\ref{qcmngtz}) with equal pair and quartet amplitudes $x=y$; ``$\Gamma_x [\Gamma_x \Gamma_y]_0$" for the state $\Gamma^\dagger_{1}(x) [\Gamma^\dagger(x)\Gamma^\dagger(y)]^{T=0} | 0 \rangle$; ``$P_{NT}|BCS\rangle_{1,2}$" for the projected BCS states of Eqs. (\ref{pbcs1},\ref{pbcs2}).}
\label{fig1}

\end{figure*}

In the QCM quartetting approach, the states for $N>Z$ systems are constructed  by appending to the $N=Z$ ansatz additional coherent pairs \cite{San12}. A state with $n_p$ excess  neutron pairs and $n_q$ quartets, having $T=T_z=n_p$ is defined as the particular combination 
\beq
\label{qcmngtz}
|QCM(T=T_z=n_p)\rangle=[\Gamma^\dagger_1(y)]^{n_p}\,[ Q^\dagger(x)]^{n_q}\,|0\rangle~.
\eeq 

Here, one allows the extra collective pairs $\Gamma^\dagger_1(y)$ to have a different structure than the pairs $\Gamma^\dagger(x)$  forming the quartets. The same idea may be applied to the BCS ansatz: below, we consider the pair condensates of Eq. (\ref{bcs2}) to have different mixing amplitudes. Note that we also have to append a $\nu\nu$ pair condensate to the $\pi\nu$ condensate in this $N>Z$ case. In this section, we define $|BCS_{II}\rangle=\exp[\Gamma^\dagger_{1}(y)] \exp[\Gamma^\dagger_{0}(x)]\,|0\rangle$. We consider as illustrative examples an $N=4,Z=2$ system and an $N=6,Z=2$ system. The particle number and isospin projected combinations are
\begin{subequations}

\begin{equation}\label{pbcs1}
\mathcal{P}^{\mathcal{N}=6}_{T=T_z=1}|BCS_I\rangle=(\Gamma^\dagger_{1,x}\, Q^\dagger_y -3 \Gamma^\dagger_{1,y}\, [\Gamma^\dagger_x\Gamma^\dagger_y]^{T=0})\, |0\rangle~,\\
\end{equation}
\begin{equation}\label{pbcs2}
\mathcal{P}^{\mathcal{N}=6}_{{T=T_z=1}}|BCS_{II}\rangle=(2\Gamma^\dagger_{1,y}\, Q^\dagger_x - \Gamma^\dagger_{1,x}\, [\Gamma^\dagger_x\Gamma^\dagger_y]^{T=0})\, |0\rangle~,\\
\end{equation}
\begin{equation}
    \begin{aligned}
    &\mathcal{P}^{\mathcal{N}=8}_{{T=T_z=2}}|BCS_I\rangle=\\
&(5\, [\Gamma^\dagger_{1,y}]^2\, [\Gamma^\dagger_y\Gamma^\dagger_x]^{T=0} -2 \,\Gamma^\dagger_{1,y}\, \Gamma^\dagger_{1,x}\, Q^\dagger_y )\, |0\rangle,\\
    \end{aligned}
\end{equation}
\begin{equation}
    \begin{aligned}
&\mathcal{P}^{\mathcal{N}=8}_{{T=T_z=2}}|BCS_{II}\rangle=(11\, [\Gamma^\dagger_{1,y}]^2\, Q^\dagger_x\\
&+4\, [\Gamma^\dagger_{1,x}]^2\, Q^\dagger_{y} - 12\, \Gamma^\dagger_{1,x}\,\Gamma^\dagger_{1,y}\, [\Gamma^\dagger_x\Gamma^\dagger_y]^{T=0})\, |0\rangle~,\\
    \end{aligned}
\end{equation}
\end{subequations}
with the notation $\Gamma^\dagger_x=\Gamma^\dagger(x)$, $Q^\dagger_y=Q^\dagger(y)$ etc.
Naturally, there are multiple options of coupling various pairs to a given total isospin, and the QCM ansatz of Eq. (\ref{qcmngtz}) is just a particular choice. Interestingly, the QCM choice does not appear in all previous expressions.

{With the states (26), we performed variation-after-projection calculations for a picket-fence model of eight doubly degenerate levels, of single particle energies $\epsilon_k=k-1$, and with a state independent interaction of strength $G$}. The analytical expressions for the average of the isovector pairing Hamiltonian on {the states (26)} were derived with the Cadabra2 computer algebra system \cite{cadabra} using the method presented in Refs.  \cite{Bar191,Bar192}. 

In all cases, we obtained a very good agreement between the projected BCS and the QCM results. For the chosen model, the overlaps do not decrease lower than $0.999$, and the relative errors in the correlation energies do not exceed 0.5\%. We present in Fig. 1 the results for the lightest $N=4, Z=2$ system; the agreement between projected BCS and QCM improves for heavier systems (we note that the QCM ansatz gives a higher correlation energy in all cases).

Note that even in the case of equal pair and quartet mixing amplitudes ($x=y)$ the results are still good: the obtained overlaps with the QCM state (having $x\neq y$) are always greater than $0.98$, and the errors in the correlation energies are always smaller than $8\%$. In this case, all analytical expressions for the projected BCS states reduce to the QCM ansatz of Eq. (\ref{qcmngtz}) with $x=y$.

In constructing the QCM ansatz for $N>Z$ systems, Ref. \cite{San12} mentions the necessity of a different structure for the excess collective neutron pairs with respect to the collective pairs forming the quartet, as to reproduce the Hartree-Fock limit. However, the present results indicate that while the $x=y$ choice introduces significant errors,
it preserves the correct behaviour in the weak pairing regime. Indeed, the Hartree-Fock vacuum may be obtained as a limit of the $x=y$ QCM ansatz by suitably scaling the mixing amplitudes. For the $N=4,Z=2$ and $N=6,Z=2$ systems with the scalings  $w_\varepsilon=(1/{\varepsilon},\varepsilon^2,0,0,\dots)$ and $z_\varepsilon=(1/{\varepsilon},\varepsilon,\varepsilon,0,0,\dots)$, we obtain
\beq
\begin{aligned}
\Gamma_1^\dagger(w_\epsilon)\, Q^\dagger(w_\varepsilon)&\sim q_1^\dagger P^\dagger_{2,1}+\mathcal{O}(\varepsilon)~,\\
[\Gamma_1^\dagger(z_\varepsilon)]^2\, Q^\dagger(z_\varepsilon)&\sim q_1^\dagger P^\dagger_{2,1}P^\dagger_{3,1}+\mathcal{O}(\varepsilon)~.\\
\end{aligned}
\eeq
which  reduce to the exact Hartree-Fock state in the  $\varepsilon\rightarrow 0$  limit.

\subparagraph{Summary and Conclusions.}

We presented an attempt at bridging the descriptions of the proton-neutron isovector pairing correlations in the symmetry preserving quartet picture and in the mean-field pair-condensate picture.

 For both the coherent and the projected state, the nonlinear annihilation operators are shown to generate nontrivial connections between the many-body correlation functions. A {possible} application of these relations would be to consider the novel quantum many-body bootstrap approach \cite{qmb_boot, boot_prl} and to implement the condensate property of the ansatz in terms of these constraints for the correlation functions. This would enable {a numerically unified description, based on a quartet coherent state, of both nuclear matter and finite nuclei.} The same framework could be generalized to quartetting in condensed matter systems e.g., to the study of bi-exciton condensation in semiconductors or trapped fermionic atoms in optical lattices.

Then, inspired by the structure of the quartet coherent state, we have shown that the QCM ansatz for the ground state of even-even $N=Z$ systems can be obtained by projecting out the particle number and the isospin from  a proton-neutron BCS state. For the $N>Z$
systems the $P_{NT}$BCS and QCM states are  not analytically equivalent. However, their overlaps are very close to one.  The numerical $P_{NT}$BCS calculations indicate that the particular way of coupling various pairs to the total isospin of the $N>Z$ system does not influence much the final results as long as the trial states obeys the correct 
symmetry constraints. An interesting question is whether these facts hold in the case of an isovector-isoscalar 
pairing Hamiltonian. This issue we intend to address in a future study.

\begin{acknowledgments}
This work was supported by a grant of the Romanian Ministry of Education and Research, CNCS - UEFISCDI,
project number PN-III-P1-1.1-PD-2019-0346, within PNCDI III, and PN-19060101/2019-2022.
\end{acknowledgments}

\bibliography{mybibfile}

\begin{thebibliography}{33}
\expandafter\ifx\csname natexlab\endcsname\relax\def\natexlab#1{#1}\fi
\expandafter\ifx\csname bibnamefont\endcsname\relax
  \def\bibnamefont#1{#1}\fi
\expandafter\ifx\csname bibfnamefont\endcsname\relax
  \def\bibfnamefont#1{#1}\fi
\expandafter\ifx\csname citenamefont\endcsname\relax
  \def\citenamefont#1{#1}\fi
\expandafter\ifx\csname url\endcsname\relax
  \def\url#1{\texttt{#1}}\fi
\expandafter\ifx\csname urlprefix\endcsname\relax\def\urlprefix{URL }\fi
\providecommand{\bibinfo}[2]{#2}
\providecommand{\eprint}[2][]{\url{#2}}

\bibitem[{\citenamefont{Bohr et~al.}(1958)\citenamefont{Bohr, Mottelson, and
  Pines}}]{Bohr58}
\bibinfo{author}{\bibfnamefont{A.}~\bibnamefont{Bohr}},
  \bibinfo{author}{\bibfnamefont{B.~R.} \bibnamefont{Mottelson}},
  \bibnamefont{and} \bibinfo{author}{\bibfnamefont{D.}~\bibnamefont{Pines}},
  \bibinfo{journal}{Phys. Rev.} \textbf{\bibinfo{volume}{110}},
  \bibinfo{pages}{936} (\bibinfo{year}{1958}),
  \urlprefix\url{https://link.aps.org/doi/10.1103/PhysRev.110.936}.

\bibitem[{\citenamefont{Frauendorf and Macchiavelli}(2014)}]{Fra14}
\bibinfo{author}{\bibfnamefont{S.}~\bibnamefont{Frauendorf}} \bibnamefont{and}
  \bibinfo{author}{\bibfnamefont{A.}~\bibnamefont{Macchiavelli}},
  \bibinfo{journal}{Progress in Particle and Nuclear Physics}
  \textbf{\bibinfo{volume}{78}}, \bibinfo{pages}{24 } (\bibinfo{year}{2014}),
  ISSN \bibinfo{issn}{0146-6410},
  \urlprefix\url{http://www.sciencedirect.com/science/article/pii/S0146641014000465}.

\bibitem[{\citenamefont{Belyaev and Solovev}(1960)}]{Bel60}
\bibinfo{author}{\bibfnamefont{B.~N.} \bibnamefont{Belyaev},
  \bibfnamefont{V.~B.~Zacharev}} \bibnamefont{and}
  \bibinfo{author}{\bibfnamefont{V.~G.} \bibnamefont{Solovev}},
  \bibinfo{journal}{J. Exptl. Theoret. Phys.} \textbf{\bibinfo{volume}{38}},
  \bibinfo{pages}{952} (\bibinfo{year}{1960}).

\bibitem[{\citenamefont{Goodman}(1979)}]{Goo79}
\bibinfo{author}{\bibfnamefont{A.~L.} \bibnamefont{Goodman}},
  \bibinfo{journal}{Adv. Nuc. Phys.} \textbf{\bibinfo{volume}{11}},
  \bibinfo{pages}{263} (\bibinfo{year}{1979}).

\bibitem[{\citenamefont{Soloviev}(1960)}]{Sol60}
\bibinfo{author}{\bibfnamefont{V.~G.} \bibnamefont{Soloviev}},
  \bibinfo{journal}{Nucl. Phys.} \textbf{\bibinfo{volume}{18}},
  \bibinfo{pages}{161} (\bibinfo{year}{1960}).

\bibitem[{\citenamefont{Br\'emond and Valatin}(1963)}]{Bre63}
\bibinfo{author}{\bibfnamefont{B.}~\bibnamefont{Br\'emond}} \bibnamefont{and}
  \bibinfo{author}{\bibfnamefont{J.}~\bibnamefont{Valatin}},
  \bibinfo{journal}{Nuclear Physics} \textbf{\bibinfo{volume}{41}},
  \bibinfo{pages}{640 } (\bibinfo{year}{1963}), ISSN \bibinfo{issn}{0029-5582},
  \urlprefix\url{http://www.sciencedirect.com/science/article/pii/0029558263905431}.

\bibitem[{\citenamefont{Flowers and Vujicic}(1963)}]{Flo63}
\bibinfo{author}{\bibfnamefont{B.}~\bibnamefont{Flowers}} \bibnamefont{and}
  \bibinfo{author}{\bibfnamefont{M.}~\bibnamefont{Vujicic}},
  \bibinfo{journal}{Nuclear Physics} \textbf{\bibinfo{volume}{49}},
  \bibinfo{pages}{586} (\bibinfo{year}{1963}), ISSN \bibinfo{issn}{0029-5582},
  \urlprefix\url{http://www.sciencedirect.com/science/article/pii/0029558263901238}.

\bibitem[{\citenamefont{Dobes and Pittel}(1998)}]{Dob98}
\bibinfo{author}{\bibfnamefont{J.}~\bibnamefont{Dobes}} \bibnamefont{and}
  \bibinfo{author}{\bibfnamefont{S.}~\bibnamefont{Pittel}},
  \bibinfo{journal}{Phys. Rev. C} \textbf{\bibinfo{volume}{57}},
  \bibinfo{pages}{688} (\bibinfo{year}{1998}),
  \urlprefix\url{https://link.aps.org/doi/10.1103/PhysRevC.57.688}.

\bibitem[{\citenamefont{Sandulescu
  et~al.}(2012{\natexlab{a}})\citenamefont{Sandulescu, Negrea, and
  Johnson}}]{San12}
\bibinfo{author}{\bibfnamefont{N.}~\bibnamefont{Sandulescu}},
  \bibinfo{author}{\bibfnamefont{D.}~\bibnamefont{Negrea}}, \bibnamefont{and}
  \bibinfo{author}{\bibfnamefont{C.~W.} \bibnamefont{Johnson}},
  \bibinfo{journal}{Phys. Rev. C} \textbf{\bibinfo{volume}{86}},
  \bibinfo{pages}{041302(R)} (\bibinfo{year}{2012}{\natexlab{a}}),
  \urlprefix\url{https://link.aps.org/doi/10.1103/PhysRevC.86.041302}.

\bibitem[{\citenamefont{Sandulescu
  et~al.}(2012{\natexlab{b}})\citenamefont{Sandulescu, Negrea, Dukelsky, and
  Johnson}}]{San12a}
\bibinfo{author}{\bibfnamefont{N.}~\bibnamefont{Sandulescu}},
  \bibinfo{author}{\bibfnamefont{D.}~\bibnamefont{Negrea}},
  \bibinfo{author}{\bibfnamefont{J.}~\bibnamefont{Dukelsky}}, \bibnamefont{and}
  \bibinfo{author}{\bibfnamefont{C.~W.} \bibnamefont{Johnson}},
  \bibinfo{journal}{Phys. Rev. C} \textbf{\bibinfo{volume}{85}},
  \bibinfo{pages}{061303(R)} (\bibinfo{year}{2012}{\natexlab{b}}),
  \urlprefix\url{https://link.aps.org/doi/10.1103/PhysRevC.85.061303}.

\bibitem[{\citenamefont{Sambataro and
  Sandulescu}(2020{\natexlab{a}})}]{Samb20a}
\bibinfo{author}{\bibfnamefont{M.}~\bibnamefont{Sambataro}} \bibnamefont{and}
  \bibinfo{author}{\bibfnamefont{N.}~\bibnamefont{Sandulescu}},
  \bibinfo{journal}{J. Phys. G: Nucl. Part. Phys.}
  \textbf{\bibinfo{volume}{47}}, \bibinfo{pages}{045112}
  (\bibinfo{year}{2020}{\natexlab{a}}),
  \urlprefix\url{https://iopscience.iop.org/article/10.1088/1361-6471/ab6ee2}.

\bibitem[{\citenamefont{Sambataro and
  Sandulescu}(2020{\natexlab{b}})}]{Samb20b}
\bibinfo{author}{\bibfnamefont{M.}~\bibnamefont{Sambataro}} \bibnamefont{and}
  \bibinfo{author}{\bibfnamefont{N.}~\bibnamefont{Sandulescu}},
  \bibinfo{journal}{J. Phys. G: Nucl. Part. Phys., in press}
  (\bibinfo{year}{2020}{\natexlab{b}}),
  \urlprefix\url{https://iopscience.iop.org/article/10.1088/1361-6471/abafff}.

\bibitem[{\citenamefont{Sogo et~al.}(2009)\citenamefont{Sogo, Lazauskas,
  R\"opke, and Schuck}}]{Sogo09}
\bibinfo{author}{\bibfnamefont{T.}~\bibnamefont{Sogo}},
  \bibinfo{author}{\bibfnamefont{R.}~\bibnamefont{Lazauskas}},
  \bibinfo{author}{\bibfnamefont{G.}~\bibnamefont{R\"opke}}, \bibnamefont{and}
  \bibinfo{author}{\bibfnamefont{P.}~\bibnamefont{Schuck}},
  \bibinfo{journal}{Phys. Rev. C} \textbf{\bibinfo{volume}{79}},
  \bibinfo{pages}{051301(R)} (\bibinfo{year}{2009}),
  \urlprefix\url{https://link.aps.org/doi/10.1103/PhysRevC.79.051301}.

\bibitem[{\citenamefont{Tohsaki et~al.}(2001)\citenamefont{Tohsaki, Horiuchi,
  Schuck, and R\"opke}}]{THSR}
\bibinfo{author}{\bibfnamefont{A.}~\bibnamefont{Tohsaki}},
  \bibinfo{author}{\bibfnamefont{H.}~\bibnamefont{Horiuchi}},
  \bibinfo{author}{\bibfnamefont{P.}~\bibnamefont{Schuck}}, \bibnamefont{and}
  \bibinfo{author}{\bibfnamefont{G.}~\bibnamefont{R\"opke}},
  \bibinfo{journal}{Phys. Rev. Lett.} \textbf{\bibinfo{volume}{87}},
  \bibinfo{pages}{192501} (\bibinfo{year}{2001}),
  \urlprefix\url{https://link.aps.org/doi/10.1103/PhysRevLett.87.192501}.

\bibitem[{\citenamefont{Schuck et~al.}(2014)\citenamefont{Schuck, Funaki,
  Horiuchi, Röpke, Tohsaki, and Yamada}}]{Schuck2014}
\bibinfo{author}{\bibfnamefont{P.}~\bibnamefont{Schuck}},
  \bibinfo{author}{\bibfnamefont{Y.}~\bibnamefont{Funaki}},
  \bibinfo{author}{\bibfnamefont{H.}~\bibnamefont{Horiuchi}},
  \bibinfo{author}{\bibfnamefont{G.}~\bibnamefont{Röpke}},
  \bibinfo{author}{\bibfnamefont{A.}~\bibnamefont{Tohsaki}}, \bibnamefont{and}
  \bibinfo{author}{\bibfnamefont{T.}~\bibnamefont{Yamada}},
  \bibinfo{journal}{Journal of Physics: Conference Series}
  \textbf{\bibinfo{volume}{529}}, \bibinfo{pages}{012014}
  (\bibinfo{year}{2014}),
  \urlprefix\url{https://doi.org/10.1088%2F1742-6596%2F529%2F1%2F012014}.

\bibitem[{\citenamefont{Baran and Delion}(2020)}]{QBCS}
\bibinfo{author}{\bibfnamefont{V.~V.} \bibnamefont{Baran}} \bibnamefont{and}
  \bibinfo{author}{\bibfnamefont{D.~S.} \bibnamefont{Delion}},
  \bibinfo{journal}{Phys. Lett. B} \textbf{\bibinfo{volume}{805}},
  \bibinfo{pages}{135462} (\bibinfo{year}{2020}), ISSN
  \bibinfo{issn}{0370-2693},
  \urlprefix\url{http://www.sciencedirect.com/science/article/pii/S0370269320302665}.

\bibitem[{\citenamefont{Sandulescu et~al.}(2009)\citenamefont{Sandulescu,
  Errea, and Dukelsky}}]{San09}
\bibinfo{author}{\bibfnamefont{N.}~\bibnamefont{Sandulescu}},
  \bibinfo{author}{\bibfnamefont{B.}~\bibnamefont{Errea}}, \bibnamefont{and}
  \bibinfo{author}{\bibfnamefont{J.}~\bibnamefont{Dukelsky}},
  \bibinfo{journal}{Phys. Rev. C} \textbf{\bibinfo{volume}{80}},
  \bibinfo{pages}{044335} (\bibinfo{year}{2009}),
  \urlprefix\url{https://link.aps.org/doi/10.1103/PhysRevC.80.044335}.

\bibitem[{\citenamefont{Romero et~al.}(2019)\citenamefont{Romero, Dobaczewski,
  and Pastore}}]{Dob19}
\bibinfo{author}{\bibfnamefont{A.}~\bibnamefont{Romero}},
  \bibinfo{author}{\bibfnamefont{J.}~\bibnamefont{Dobaczewski}},
  \bibnamefont{and} \bibinfo{author}{\bibfnamefont{A.}~\bibnamefont{Pastore}},
  \bibinfo{journal}{Physics Letters B} \textbf{\bibinfo{volume}{795}},
  \bibinfo{pages}{177 } (\bibinfo{year}{2019}), ISSN \bibinfo{issn}{0370-2693},
  \urlprefix\url{http://www.sciencedirect.com/science/article/pii/S0370269319304113}.

\bibitem[{\citenamefont{Schuck}(2020)}]{Schuck2020}
\bibinfo{author}{\bibfnamefont{P.}~\bibnamefont{Schuck}},
  \bibinfo{journal}{International Journal of Modern Physics E}
  \textbf{\bibinfo{volume}{29}}, \bibinfo{pages}{2050023}
  (\bibinfo{year}{2020}), \eprint{https://doi.org/10.1142/S0218301320500238},
  \urlprefix\url{https://doi.org/10.1142/S0218301320500238}.

\bibitem[{\citenamefont{Kyotoku and Chen}(1979)}]{Kyotoku79}
\bibinfo{author}{\bibfnamefont{M.}~\bibnamefont{Kyotoku}} \bibnamefont{and}
  \bibinfo{author}{\bibfnamefont{H.-T.} \bibnamefont{Chen}},
  \bibinfo{journal}{Journal of Physics G: Nuclear Physics}
  \textbf{\bibinfo{volume}{5}}, \bibinfo{pages}{1649} (\bibinfo{year}{1979}),
  \urlprefix\url{https://doi.org/10.1088%2F0305-4616%2F5%2F12%2F007}.

\bibitem[{\citenamefont{Kyotoku and Chen}(1987)}]{Kyotoku87}
\bibinfo{author}{\bibfnamefont{M.}~\bibnamefont{Kyotoku}} \bibnamefont{and}
  \bibinfo{author}{\bibfnamefont{H.-T.} \bibnamefont{Chen}},
  \bibinfo{journal}{Phys. Rev. C} \textbf{\bibinfo{volume}{36}},
  \bibinfo{pages}{1144} (\bibinfo{year}{1987}),
  \urlprefix\url{https://link.aps.org/doi/10.1103/PhysRevC.36.1144}.

\bibitem[{\citenamefont{Chen et~al.}(1978)\citenamefont{Chen, Müther, and
  Faessler}}]{Chen78}
\bibinfo{author}{\bibfnamefont{H.-T.} \bibnamefont{Chen}},
  \bibinfo{author}{\bibfnamefont{H.}~\bibnamefont{Müther}}, \bibnamefont{and}
  \bibinfo{author}{\bibfnamefont{A.}~\bibnamefont{Faessler}},
  \bibinfo{journal}{Nuclear Physics A} \textbf{\bibinfo{volume}{297}},
  \bibinfo{pages}{445 } (\bibinfo{year}{1978}), ISSN \bibinfo{issn}{0375-9474},
  \urlprefix\url{http://www.sciencedirect.com/science/article/pii/0375947478901549}.

\bibitem[{\citenamefont{Raduta and {Moya de Guerra}}(2000)}]{Raduta00}
\bibinfo{author}{\bibfnamefont{A.~A.} \bibnamefont{Raduta}} \bibnamefont{and}
  \bibinfo{author}{\bibfnamefont{E.}~\bibnamefont{{Moya de Guerra}}},
  \bibinfo{journal}{Annals of Physics} \textbf{\bibinfo{volume}{284}},
  \bibinfo{pages}{134 } (\bibinfo{year}{2000}), ISSN \bibinfo{issn}{0003-4916},
  \urlprefix\url{http://www.sciencedirect.com/science/article/pii/S0003491600960652}.

\bibitem[{\citenamefont{Raduta et~al.}(2001)\citenamefont{Raduta, Sarriguren,
  Faessler, and {de Guerra}}}]{Raduta01}
\bibinfo{author}{\bibfnamefont{A.~A.} \bibnamefont{Raduta}},
  \bibinfo{author}{\bibfnamefont{P.}~\bibnamefont{Sarriguren}},
  \bibinfo{author}{\bibfnamefont{A.}~\bibnamefont{Faessler}}, \bibnamefont{and}
  \bibinfo{author}{\bibfnamefont{E.}~\bibnamefont{{de Guerra}}},
  \bibinfo{journal}{Annals of Physics} \textbf{\bibinfo{volume}{294}},
  \bibinfo{pages}{182 } (\bibinfo{year}{2001}), ISSN \bibinfo{issn}{0003-4916},
  \urlprefix\url{http://www.sciencedirect.com/science/article/pii/S0003491601961834}.

\bibitem[{\citenamefont{Raduta et~al.}(2000)\citenamefont{Raduta, Pacearescu,
  Baran, Sarriguren, and {Moya de Guerra}}}]{Raduta00a}
\bibinfo{author}{\bibfnamefont{A.~A.} \bibnamefont{Raduta}},
  \bibinfo{author}{\bibfnamefont{L.}~\bibnamefont{Pacearescu}},
  \bibinfo{author}{\bibfnamefont{V.}~\bibnamefont{Baran}},
  \bibinfo{author}{\bibfnamefont{P.}~\bibnamefont{Sarriguren}},
  \bibnamefont{and} \bibinfo{author}{\bibfnamefont{E.}~\bibnamefont{{Moya de
  Guerra}}}, \bibinfo{journal}{Nuclear Physics A}
  \textbf{\bibinfo{volume}{675}}, \bibinfo{pages}{503 } (\bibinfo{year}{2000}),
  ISSN \bibinfo{issn}{0375-9474},
  \urlprefix\url{http://www.sciencedirect.com/science/article/pii/S0375947400001834}.

\bibitem[{\citenamefont{Raduta et~al.}(2012)\citenamefont{Raduta,
  Krivoruchenko, and Faessler}}]{Raduta12}
\bibinfo{author}{\bibfnamefont{A.~A.} \bibnamefont{Raduta}},
  \bibinfo{author}{\bibfnamefont{M.~I.} \bibnamefont{Krivoruchenko}},
  \bibnamefont{and} \bibinfo{author}{\bibfnamefont{A.}~\bibnamefont{Faessler}},
  \bibinfo{journal}{Phys. Rev. C} \textbf{\bibinfo{volume}{85}},
  \bibinfo{pages}{054314} (\bibinfo{year}{2012}),
  \urlprefix\url{https://link.aps.org/doi/10.1103/PhysRevC.85.054314}.

\bibitem[{\citenamefont{Ring and Schuck}(1980)}]{RS}
\bibinfo{author}{\bibfnamefont{P.}~\bibnamefont{Ring}} \bibnamefont{and}
  \bibinfo{author}{\bibfnamefont{P.}~\bibnamefont{Schuck}},
  \emph{\bibinfo{title}{The Nuclear Many-Body Problem}}
  (\bibinfo{publisher}{Springer}, \bibinfo{year}{1980}).

\bibitem[{\citenamefont{Baran and Delion}(2019{\natexlab{a}})}]{phb}
\bibinfo{author}{\bibfnamefont{V.~V.} \bibnamefont{Baran}} \bibnamefont{and}
  \bibinfo{author}{\bibfnamefont{D.~S.} \bibnamefont{Delion}},
  \bibinfo{journal}{Phys. Rev. C} \textbf{\bibinfo{volume}{99}},
  \bibinfo{pages}{064311} (\bibinfo{year}{2019}{\natexlab{a}}),
  \urlprefix\url{https://link.aps.org/doi/10.1103/PhysRevC.99.064311}.

\bibitem[{\citenamefont{Han}(2020)}]{qmb_boot}
\bibinfo{author}{\bibfnamefont{X.}~\bibnamefont{Han}} (\bibinfo{year}{2020}),
  \eprint{arXiv: 2006.06002}.

\bibitem[{\citenamefont{Han et~al.}(2020)\citenamefont{Han, Hartnoll, and
  Kruthoff}}]{boot_prl}
\bibinfo{author}{\bibfnamefont{X.}~\bibnamefont{Han}},
  \bibinfo{author}{\bibfnamefont{S.~A.} \bibnamefont{Hartnoll}},
  \bibnamefont{and} \bibinfo{author}{\bibfnamefont{J.}~\bibnamefont{Kruthoff}},
  \bibinfo{journal}{Phys. Rev. Lett.} \textbf{\bibinfo{volume}{125}},
  \bibinfo{pages}{041601} (\bibinfo{year}{2020}),
  \urlprefix\url{https://link.aps.org/doi/10.1103/PhysRevLett.125.041601}.

\bibitem[{\citenamefont{Peeters}(2018)}]{cadabra}
\bibinfo{author}{\bibfnamefont{K.}~\bibnamefont{Peeters}},
  \bibinfo{journal}{Journal of Open Source Software}
  \textbf{\bibinfo{volume}{3(32)}}, \bibinfo{pages}{1118}
  (\bibinfo{year}{2018}), \urlprefix\url{https://doi.org/10.21105/joss.01118}.

\bibitem[{\citenamefont{Baran and Delion}(2019{\natexlab{b}})}]{Bar191}
\bibinfo{author}{\bibfnamefont{V.~V.} \bibnamefont{Baran}} \bibnamefont{and}
  \bibinfo{author}{\bibfnamefont{D.~S.} \bibnamefont{Delion}},
  \bibinfo{journal}{Phys. Rev. C} \textbf{\bibinfo{volume}{99}},
  \bibinfo{pages}{031303} (\bibinfo{year}{2019}{\natexlab{b}}),
  \urlprefix\url{https://link.aps.org/doi/10.1103/PhysRevC.99.031303}.

\bibitem[{\citenamefont{Baran et~al.}(2019)\citenamefont{Baran, Delion, and
  Dolteanu}}]{Bar192}
\bibinfo{author}{\bibfnamefont{V.~V.} \bibnamefont{Baran}},
  \bibinfo{author}{\bibfnamefont{D.~S.} \bibnamefont{Delion}},
  \bibnamefont{and} \bibinfo{author}{\bibfnamefont{S.}~\bibnamefont{Dolteanu}},
  \bibinfo{journal}{Phys. Rev. C} \textbf{\bibinfo{volume}{100}},
  \bibinfo{pages}{034326} (\bibinfo{year}{2019}),
  \urlprefix\url{https://link.aps.org/doi/10.1103/PhysRevC.100.034326}.

\end{thebibliography}

\end{document}